\documentclass[sigconf]{acmart}
\usepackage{cleveref} 
\usepackage{multirow}
\usepackage{booktabs}
\usepackage{enumitem}
\usepackage{graphicx}
\usepackage{array}
\usepackage{caption}  

\usepackage{array,makecell,tabularx}
\newcolumntype{C}[1]{>{\centering\arraybackslash}m{#1}} 
\newcolumntype{Y}{>{\raggedright\arraybackslash}X}      

\definecolor{yipeng_color}{RGB}{250, 50, 50}

\AtBeginDocument{%
  }

\setcopyright{acmlicensed}
\copyrightyear{2025}
\acmYear{2025}
\acmDOI{XXXXXXX.XXXXXXX}

\acmISBN{978-1-4503-XXXX-X/18/06}

\begin{document}

\title{Bridging Modality Gaps in e-Commerce Products via Vision-Language Alignment}

\author{Yipeng Zhang}
\authornotemark[1]
\email{yipezhang@ebay.com}
\thanks{Equal contribution.}
\affiliation{%
  \institution{eBay Inc.}
  \city{San Jose}
  \state{CA}
  \country{USA}
}

\author{Hongjun Yu}
\authornotemark[1]
\email{hongjyu@ebay.com}
\affiliation{%
  \institution{eBay Inc.}
    \city{San Jose}
  \state{CA}
  \country{USA}
}

\author{Aritra Mandal}
\authornotemark[1]
\email{arimandal@ebay.com}
\affiliation{%
  \institution{eBay Inc.}
    \city{San Jose}
  \state{CA}
  \country{USA}
}

\author{Canran Xu}
\authornotemark[1]
\email{canxu@ebay.com}
\affiliation{%
  \institution{eBay Inc.}
  \city{Shanghai}
  \country{China}
}

\author{Qunzhi Zhou}
\authornotemark[1]
\email{qunzhou@ebay.com}
\affiliation{%
  \institution{eBay Inc.}
    \city{San Jose}
  \state{CA}
  \country{USA}
}

\author{Zhe Wu}
\authornotemark[1]
\email{zwu1@ebay.com}
\affiliation{%
  \institution{eBay Inc.}
  \city{San Jose}
  \state{CA}
  \country{USA}
}

\renewcommand{\shortauthors}{Zhang et al.}


\begin{abstract}
Item information, such as titles and attributes, is essential for effective user engagement in e-commerce. However, manual or semi-manual entry of structured item specifics often results in inconsistent data quality, errors, and a time-intensive process, particularly for Customer-to-Customer sellers. Generating these descriptions directly from item images offers a promising solution. Existing retrieval-based solutions partially address these issues, but often fall short in capturing fine-grained visual details and handling niche or specialized product categories. To address these challenges, we propose \textbf{O}ptimized \textbf{P}reference-Based \textbf{A}I for \textbf{L}istings (OPAL), a novel framework to directly generate schema-compliant, high-quality item descriptions from images using a fine-tuned multimodal-large language model (MLLM). OPAL addresses key challenges in multimodal learning in e-commerce, such as modality gaps and the need for fine-grained contextual understanding. Specifically, OPAL integrates novel data refinement methods—MLLM-Assisted Conformity Enhancement and LLM-Assisted Contextual Understanding—to align textual and visual information while addressing the granularity of contextual understanding. We leverage visual instruction tuning and direct preference optimization to fine-tune MLLM, mitigating hallucination risks and improving model performance across different backbones. Through extensive experiments on real-world e-commerce datasets, we demonstrate that OPAL consistently outperforms baseline solutions in generation quality and completion rates. These results highlight OPAL's effectiveness in bridging the modality gap and elevating the standard of automated listing optimization in e-commerce.

\end{abstract}




\maketitle

\begin{figure}[!h]
    \centering
    \includegraphics[width=0.35\textwidth]{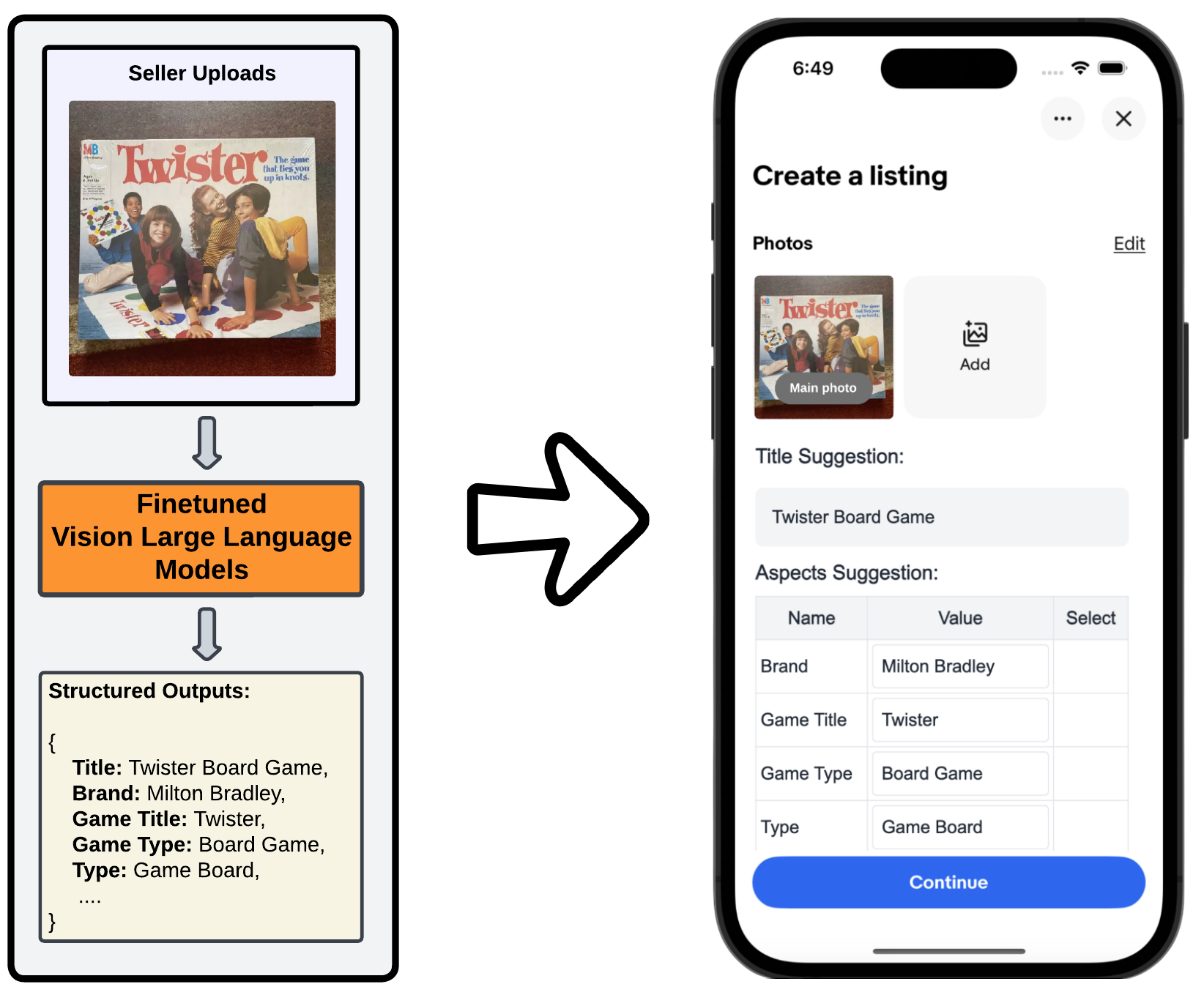}
    \caption{Overview of the proposed use case. When a user uploads an image, OPAL generates a structured output including title and aspect name-value pairs. This structured output is pre-filled for the user, streamlining the listing process and ensuring schema-compliant, high-quality descriptions.}
    \label{fig:fig1}
\end{figure}

\section{Introduction}

In e-commerce, structured item information are essential for effective search, navigation, and user experience enhancements. These descriptions, comprising an informative \textbf{title} and a series of \textbf{aspect name-value pairs} (e.g., "Department: Men," "Brand: Nike") are typically provided by sellers during the listing process. However, manual or semi-manual entry often results in inconsistent data quality, with errors, omissions, and irrelevant information. This process is also time-intensive, especially for Customer-to-Customer (C2C) sellers, dissuading them from listing more items. However, the images of items are rich in information and intent, using the images to infer obvious information aids the seller in filling relevant information faster and with more reliability. Using images to populate item information makes the listing creation more of a review process than a data entry process.

Generating structured descriptions from item images offers a promising solution, but faces significant challenges, especially in C2C scenarios with informal photography, cluttered backgrounds, and varied shooting styles. Although advanced vision models such as YOLO~\cite{wang2024yolov10} and SAM~\cite{Kirillov_2023_ICCV} excel at object detection, they fail to capture the seller's intent or complex attribute relationships. Retrieval-based methods, such as CLIP~\cite{radford2021learning}, partially address this by matching images to predefined product candidates, and recent multimodal large language models (MLLM) approach, IPL~\cite{Chen2024IPL} extends this by generating text based on image retrieval results. However, these inventory-dependent methods often fall short in accurately capturing fine-grained visual details and struggle with less common or specialized product categories due to sparse domain-specific information.

Rather than relying on similarity-based~\cite{mandal2023semanticequivalenceecommercequeries} retrieval systems, we propose a novel approach that fine-tunes an MLLM to directly generate structured item descriptions, as illustrated in Figure~\ref{fig:fig1}. Our method produces titles and aspects grounded in item images while addressing three critical criteria: (1) Alignment with Seller Intent (capturing the essence of what the seller wants to list), (2) Alignment with Image Information (reflecting visually inferable details such as color or brand), and (3) Conformity to E-commerce Standards (adhering to predefined schemas for aspects like "Model" or "Publisher"). The proposed method consumes images and extracts visually deductible information, infers relationships, and aligns with knowledge and ontology structure.

Training of MLLM requires high-quality image-text pairs that accurately map product details to their visual context. Although general-purpose datasets like MS COCO~\cite{lin2014microsoft}, Conceptual Captions \cite{sharma2018conceptual}, and Flickr30k \cite{plummer2015flickr30k} are well curated and annotated, e-commerce datasets, especially the product inventory, are inherently noisy and exhibit a significant modality gap. Textual descriptions often include irrelevant information, such as promotional statements, while visual content may inadequately represent key attributes (e.g., storage capacity or dimension). This misalignment increases the risk of hallucination, leading to unreliable outputs. Moreover, the level of granularity required for item recognition in e-commerce far exceeds that of public datasets. Without explicit reinforcement of contextual understanding, the generalizability of the trained MLLM remains limited, compromising its performance in real-world scenarios.

To address these challenges, we developed \textbf{O}ptimized \textbf{P}reference-Based \textbf{A}I for \textbf{L}istings (OPAL) equipped with a comprehensive training pipeline that integrates visual and textual signals to produce high quality listings compliant with the schema (Figure~\ref{fig:fig1}). Our approach mitigates the modality gap in the e-commerce data while optimizing model performance through visual instruction tuning. The model is further trained with direct preference optimization to enhance contextual understanding. Our contributions can be summarized as follows.

\begin{itemize}[leftmargin=*]
\item We analyze the key challenges of multimodal learning in e-commerce, focusing on both the modality gap and the granularity of contextual understanding.
\item We propose an effective training strategy (OPAL) to address the modality gap and enhance contextual understanding through refined data and model training. Extensive experiments demonstrate that this strategy consistently improves generation quality across different MLLM backbones.
\item We showcase OPAL’s superiority over in-house solutions, achieving significant improvements in generation quality and completion rates.
\end{itemize}

The proposed modality-aligned vision-language model is designed to power an API that extracts and infers structured item information directly from images. These structured outputs can then be applied across various e-commerce applications, such as product listing generation and search relevance enhancement.

\section{Related Works}

\noindent\paragraph{\textbf{Attribute Value Extraction.}}
Attribute value extraction is crucial in e-commerce for structuring product information from unstructured text. Traditional methods relied on rule-based systems and sequence labeling models like CRFs~\cite{Majumder2020Representation}, which demanded extensive manual effort and struggled with scalability. The introduction of transformer-based models, such as BERT~\cite{devlin2018bert}, significantly improved extraction accuracy by leveraging contextual embeddings. Frameworks like OpenTag~\cite{zheng2018opentag} introduced pointer networks to handle noisy and fragmented product text. More recently, LLM-based approaches like ExtractGPT~\cite{brinkmann2025extractgpt} have enabled zero-shot and few-shot extraction, demonstrating strong generalization to unseen attribute values. Multilingual models such as GAVEL~\cite{hongwimol2025gavel} further extend this capability to diverse markets and languages.

\noindent\paragraph{\textbf{Large Language Models (LLMs)}}
Large Language Models, including GPT-4~\cite{openai2023gpt}, PaLM~\cite{chowdhery2022palm}, and LLaMA~\cite{touvron2023llama}, have transformed natural language understanding and generation tasks. In e-commerce, LLMs have been adapted for product categorization, attribute-value generation, personalized recommendations, and listing optimization~\cite{Chen2024Personalization, Li2021Personalized}. Domain-specific LLMs like e-LLaMA~\cite{herold2025domain} improve performance by incorporating proprietary knowledge while preserving data security. Strategies involving schema alignment and user intent modeling further increase the utility of LLMs in structured prediction and personalization tasks~\cite{Zhang2024LLAShoppingAssistant, Peng2024ECeLLM, Chester2024LLM}.

\noindent\paragraph{\textbf{Multimodal Large Language Models (MLLMs).}}
MLLMs integrate visual and textual modalities to enable comprehensive multimodal reasoning. Early models such as CLIP~\cite{radford2021learning} and ALIGN~\cite{jia2022scaling} introduced contrastive learning frameworks to align image and text embeddings for robust zero-shot transfer. More advanced MLLMs like LLaVA~\cite{liu2023llava} QwenVL~\cite{wang2024qwen2} and InternVL~\cite{chen2024internvl} combine visual encoders with LLMs to enable tasks such as visual question answering, image-conditioned generation, and multimodal dialogue. These models are especially relevant in e-commerce scenarios where both product images and textual descriptions are available and complementary.

\noindent\paragraph{\textbf{MLLMs for E-commerce Applications.}}
MLLMs have increasingly been applied to e-commerce for multimodal product understanding, including title rewriting, attribute extraction, and visual-grounded content generation. For instance, models like PUMGPT \cite{xue2023pumgpt} and VL-GPT \cite{zhu2023vl} leverage visual context to improve contextual understanding. E-commerce-specific MLLMs such as IPL~\cite{Chen2024IPL} integrate product images, attributes, and category information through image retrieval of the e-commerce database. However, challenges persist due to noisy input data, weak alignment between visual and textual modalities, and inconsistency in attribute schemas~\cite{Hu2024Denoised}. Our work addresses these limitations by improving data alignment and leveraging MLLMs in a structured, schema-aware fashion.
\begin{figure}[h]
    \centering
    \includegraphics[width=0.45\textwidth]{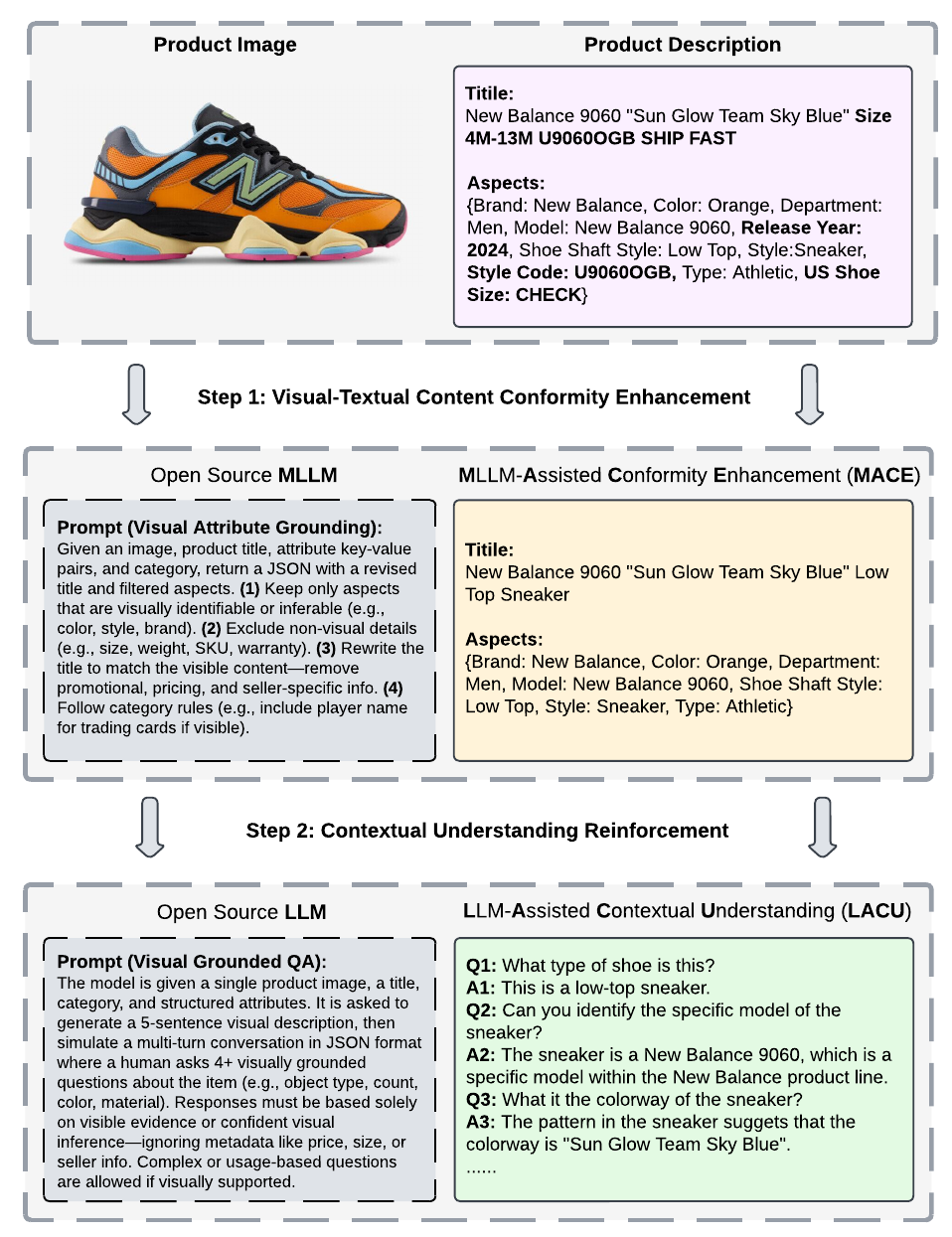}
    \caption{Overview of the modality gap in e-commerce product data and the proposed MACE–LACU pipeline. The top row illustrates the modality gap commonly observed in e-commerce listings: text descriptions often include information not visually grounded in the image, such as “Size 4M-13M” and seller-specific claims like “SHIP FAST.” Training vision-language models on such misaligned data can lead to hallucinations during inference (see Figure \ref{fig:fig3}), as visually unverifiable details could vary across listings. To mitigate this, each image–text pair is processed by MACE to filter out non-visual attributes and rewrite the product title based on image-grounded features. Then, LACU generates multi-turn, visually grounded conversations that reinforce item-specific contextual understanding. The full pipeline consumes an item image, title, category, and structured attributes, and outputs a harmonized, vision-consistent representation for improved multimodal understanding. Both prompts used in MACE and LACU are simplified due to space constraints.} 
    \label{fig:fig2}
\end{figure}

\section{Key Challenges in Multimodal Alignment and Contextual Understanding in E-commerce}

\paragraph{\textbf{Modality Gap}}
The modality gap is well-acknowledged in multimodal learning \cite{xu-etal-2021-videoclip,ModalityGap,zhang2022contrastive} and refers to the inherent differences in representation, structure, and distribution across modalities like text, image, audio, or video. This gap poses significant challenges for e-commerce datasets, where textual descriptions and image representations often lack alignment, which can result in hallucinations, where models generate plausible but incorrect information due to ambiguous or missing visual representation. Key challenges include: \textbf{1) Limitations of Visual Content:} Many product attributes are not visually discernible, such as precise specifications (e.g., “\textit{120 cm × 60 cm × 75 cm}” for furniture dimensions), materials (e.g., “\textit{100\% cotton}” for clothing), or unique identifiers (e.g., “\textit{AB1235-567}” for sneakers). Safety certifications and technical details like processor types or battery capacities also can not be represented by images solely unless with explicit textual descriptions in the image.  \textbf{2) Text-Visual Misalignment:} E-commerce listings often include irrelevant or distracting textual elements, such as promotional content (e.g., “\textit{50\% Off!}”), shipping details, or decorative emojis. These details are frequently misaligned with visual content, leading to challenges in training MLLMs. We demonstrate such modality gap using e-commerce example data in Figure~\ref {fig:fig2}.


\paragraph{\textbf{Contextual Understanding Gap in MLLM Pretraining}}
Most MLLMs are pretrained on open-source vision-language datasets such as MS-COCO~\cite{lin2014microsoft}, Conceptual Captions~\cite{sharma2018conceptual}, and Open Images~\cite{kuznetsova2020open}. While effective for general-purpose multimodal understanding, these datasets lack the contextual specificity and attribute granularity required for e-commerce applications. Their captions tend to be surface-level—e.g., “a man riding a bike” or “a red handbag on a table”—and rarely capture fine-grained product attributes critical to structured commerce systems. Examples include shoe models (e.g., “Nike Dunk Low,” “New Balance 550”), collectible brands (e.g., “Kup Stax,” “Funko Pop!,” “Bearbrick”), toy character names (e.g., “Grogu (Baby Yoda),” “Optimus Prime,” “Elsa from Frozen”), or author names in collectible books (e.g., “K.T. Oslin,” “Haruki Murakami”). These attributes are essential for item selling in e-commerce. As a result, MLLMs pretrained on generic datasets often struggle to generalize to commerce-specific tasks that demand precise visual-textual grounding and context-sensitive reasoning.

\begin{table*}[t]
\centering
\caption{Demonstration of the procedure for generating preference pairs used in DPO training. An MLLM (InternVL2.5-78B) evaluates the generated title and aspects against the visually and textually aligned item information using an evaluation prompt. If the generation is judged as \textit{Incorrect} or \textit{Mostly Incorrect}, the item information (chosen) and model output (rejected) form a preference pair for DPO training. In this example, the model incorrectly treats a golf head cover as an animal toy (highlighted). }
\renewcommand{\arraystretch}{1.2}
\footnotesize
\begin{tabular}{|m{2.5cm}|m{4.8cm}|m{3.5cm}|m{3.5cm}|m{1.5cm}|}
\hline
\textbf{Image} & \textbf{Example (Simplified) Judge Prompt} & \textbf{MACE Aligned (Chosen)} & \textbf{Model Output (Rejected)} & \textbf{Judge} \\
\hline
\centering\includegraphics[width=2.4cm, height=2.4cm, keepaspectratio]{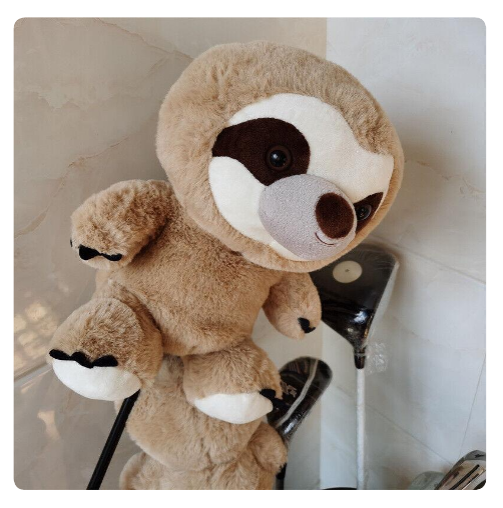} &
\scriptsize \textbf{Task}:

\scriptsize Evaluate how accurately the predicted title and aspects match the \{GOLDENTITLE\} and \{GOLDENASPECTS\}. 

\textbf{Visual Check}:

Assess visible properties in the image (color, shape, text, logos). 

\textbf{Aspect Check}: 

Match predicted aspects with visible properties and labeled aspects. 

\textbf{Title Check}: 

Check if the predicted title matches the item.

\textbf{Judgment}: 

Label \textit{each aspect} as "Correctly Identified" or "Not Correctly Identified" and the title as "Correctly Matched" or "Not Correctly Matched." 

\textbf{Final Evaluation Criteria}: 

Correctly Generated (95-100\%) Mostly Correctly Generated (80-94\%) Partially Correctly Generated (50-79\%) Mostly Incorrectly Generated (30-49\%) Incorrectly Generated (<30\%) &
\scriptsize \textbf{Title:}

Sloth Golf Wood Headcover Fit Driver Fairway Woods Club \underline{Plush Head Covers}

\quad

\textbf{Aspects:}

\{Brand :  Unbranded,  Sport/Activity :  Golf,  Type :  Wood Head Covers,  Vintage :  No\}
 &
\scriptsize \textbf{Title:}

Sloth Plush Doll Stuffed \underline{Animal Toy}

\quad

\textbf{Aspects:}

\{Brand : Unbranded, 
Character : Sloth, 
Material : Plush, 
Theme : Animation, 
Type : Action Figure
\} &
\scriptsize Incorrectly 

Generated
\\
\hline
\end{tabular}
\label{tab:dpo_prompt_example}
\end{table*}

\section{Optimized Preference-Based AI for Listings}
To address the challenges of multimodal learning in e-commerce, we propose the \textbf{O}ptimized \textbf{P}reference-Based \textbf{A}I for \textbf{L}istings (OPAL) model. OPAL integrates data enhancement, visual instruction tuning, and preference optimization to bridge modality gaps and improve contextual understanding. Its data pipeline comprises two components: \textbf{MLLM-Assisted Conformity Enhancement (MACE)} and \textbf{LLM-Assisted Contextual Understanding (LACU)}. MACE aligns visual and textual data by refining noisy e-commerce inputs, while LACU enriches context through domain-specific, multi-turn conversations. As illustrated in Figure~\ref{fig:fig2}, these methods produce well-aligned, context-rich training data. Visual instruction tuning is further improved using Direct Preference Optimization (DPO), enabling the model to generate accurate, schema-compliant titles and aspects for e-commerce listings.

\subsection{MLLM-Assisted Conformity Enhancement}

To address the modality gap in e-commerce data, we introduce MACE, a preprocessing pipeline that leverages a very large MLLM, InternVL2.5-78B, to refine the alignment between visual and textual modalities. The key idea is to enforce \textit{visual conformity} by filtering out textual information that cannot be visually confirmed or reasonably inferred from the image. Given a product image and its associated description (including title and structured aspects), the MLLM is prompted to perform two operations: (2) it rewrites the title by removing tokens that are not grounded in visual evidence; (2) it drops aspect key–value pairs that cannot be confirmed or inferred from the image. 

A simplified example of this prompting strategy is illustrated in Figure~\ref{fig:fig2}, Step 1. MACE process removes irrelevant or unverifiable details, such as shoe size, style code, or seller-specific information like ``SHIP FAST''. This results in a conformity-enhanced description that is tightly coupled with the visual content. By training downstream models on these visually grounded inputs, MACE enforces a stricter visual-to-text mapping and significantly reduces the modality gap.

\subsection{LLM-Assisted Contextual Understanding}

Although MACE effectively aligns text and image modalities, it does not fully capture the granular contextual nuances essential for e-commerce, many of which fall outside the scope of standard multimodal LLM pretraining. To address this limitation, LACU enhances MACE by generating conversational datasets using the textual LLM, LLaMA 3.1-Instruct \cite{dubey2024llama}. Inspired by common practices in visual instruction tuning \cite{liu2023llava}, we utilize this language model to simulate multi-turn dialogues that emulate interactions between a seller and a prospective buyer. These conversations are intentionally constructed to cover a broad (as many as possible) range of product aspects, thereby enriching the model’s contextual understanding. A simplified example of the prompt used to query the textual LLM is shown in Figure~\ref{fig:fig2} - Step 2. The synthetic dialogues are designed to span diverse domain-specific queries and detailed product descriptions, equipping the model to achieve a finer-grained understanding of item semantics from the image.

\subsection{Visual Instruction Tuning with DPO} We fine-tune our MLLM backbones using a visual instruction tuning on datasets derived from MACE and LACU. For each image-instruction pair, the model is optimized to minimize the loss on the generated outputs, thereby learning to produce accurate and contextually grounded product titles and attributes based on visual input. Specifically, for each (image-instruction, response) pair $(x, y)$, the model is trained to minimize the negative log-likelihood of the output sequence using standard supervised fine-tuning:

\begin{equation}
\mathcal{L}_{\text{Visual Instruction Tuning}} = - \sum_{t=1}^{T} \log P_\theta(y_t \mid y_{<t}, x)
\end{equation}

We then apply Direct Preference Optimization (DPO)~\cite{dpo} to further enhance the model's contextual understanding and reduce hallucinations. Specifically, after visual instruction tuning, we judge the model's generation on a subset of the training data using a large vision-language model, InternVL2.5-78B. This judge model receives the image, the ground truth product description, and the model-generated description, and classifies the generation into one of five categories: \textit{Correctly Generated}, \textit{Mostly Correctly Generated}, \textit{Partially Correctly Generated}, \textit{Mostly Incorrectly Generated}, or \textit{Incorrectly Generated}. Generations deemed to have incorrect contextual understanding (labeled as \textit{Mostly Incorrectly Generated} or \textit{Incorrectly Generated}) are used to construct preference pairs for DPO training, where the ground truth is treated as the \textit{chosen} output and the model prediction as the \textit{rejected} output. By creating the pair $(x, y_{\text{chosen}}, y_{\text{rejected}})$. The sigmoid-based DPO loss encourages the model to assign a higher likelihood to the preferred output:
\begin{equation}
\mathcal{L}_{\text{pref}} = -\log \sigma\left( \beta \left[ \log \pi_\theta(y_{\text{chosen}} \mid x) - \log \pi_\theta(y_{\text{rejected}} \mid x) \right] \right)
\end{equation}

A KL-regularization term is included to constrain the updated model to stay close to a reference policy:

\begin{equation}
\mathcal{L}_{\text{DPO}} = \mathcal{L}_{\text{pref}} + \lambda \cdot \mathrm{KL}\left( \pi_\theta \,\|\, \pi_{\text{ref}} \right),
\end{equation}
where $\pi_\theta$ is the current policy, $\pi_{\text{ref}}$ is the reference policy, $\sigma(\cdot)$ is the sigmoid function, $\beta$ is a temperature parameter controlling preference sharpness, and $\lambda$ weights the optional KL regularization term. A simplified illustration of this procedure, along with an example prompt, is provided in Table~\ref{tab:dpo_prompt_example}.

\section{Experiments}
\subsection{Experiment Setup}

\subsubsection{Training and Evaluation Datasets:} We collect data comprising millions of product entries from a world-leading e-commerce platform. After an extensive cleaning process, we retain one million valid image–description pairs (titles and aspects). For resource efficiency, only the main image is retained for each product.

\textbf{Task-Specific Instruction Dataset:}  Using MACE, we align the data with product images and textual description, ensuring conformity to the e-commerce schema. This process results in 890K high-quality image–instruction pairs in JSON format, each containing a refined title and corresponding aspect-value pairs.

\textbf{Contextual Understanding Dataset:}  To enhance contextual understanding, we apply LACU to generate approximately 800K image–conversation pairs. Each pair simulates multi-turn buyer-seller interactions, typically involving at least five conversational rounds. These dialogues provide a rich source of domain-specific and contextually diverse training data. Both the instruction and conversational datasets are merged during training.

\textbf{Preference Optimization Dataset:}
We use a large MLLM to evaluate each backbone's responses on a subset of the training data (200K). We then select 20K preference pairs for each backbone, each containing an instruction paired with a preferred (chosen) and a dispreferred (rejected) response.

\textbf{Evaluation Dataset:} We collect an independent dataset of 100K diverse samples comprising images and textual information (titles and aspects) from the e-commerce inventory. These samples undergo human evaluation to ensure accurate ground truth, serving as the benchmark for all downstream analysis. Notably, the dataset contains niche items that appear only after the model has been trained, providing a meaningful test of generalization.

\subsubsection{Implementation Details}

We adopt a two-stage training procedure consisting of visual instruction tuning followed by DPO. All experiments are conducted using the \texttt{LLaMA-Factory} \cite{zheng2024llamafactory} framework with full-parameter fine-tuning. The vision encoder is frozen throughout. Training is performed on 8$\times$A100 GPUs (80GB) using DeepSpeed ZeRO-3 \cite{rajbhandari2020zero} and \texttt{bfloat16} precision. For visual instruction fine-tuning, the maximum token length is set to 4,096. We use a per-device batch size of 2 with gradient accumulation over 16 steps. Models are trained for 1 epoch using AdamW with a learning rate of $1\times10^{-5}$, cosine learning rate decay, and a 10\% warmup ratio. For DPO, the pairwise preference dataset is formatted with the same prompt template. The model is optimized using a sigmoid-based DPO loss. Training is conducted for 1 epoch with a per-device batch size of 1 and gradient accumulation over 8 steps. We use a learning rate of $5\times10^{-6}$ and the same scheduler as in the previous stage.

\subsection{Quantitative Evaluation}
\begin{table}[]
\footnotesize
\centering
\caption{Performance comparison of different MLLM backbones and training strategies. An in-context learning evaluation for InternVL2.5-78B is also included for reference.} \label{tab:main_results}
\begin{tabular}{@{}ccccccc@{}}
\toprule
\multirow{2}{*}{Backbone}   &  \multicolumn{3}{c}{Strategy}   & \multirow{2}{*}{Rouge-L} & \multirow{2}{*}{Aspect} & \multirow{2}{*}{Schema} \\ 
\cmidrule(lr){2-4}
& \multicolumn{1}{c}{MACE} & \multicolumn{1}{c}{LACU} & \multicolumn{1}{c}{DPO} & \multicolumn{1}{c}{F1} &  \multicolumn{1}{c}{F1}   &     \multicolumn{1}{c}{\quad Recall}   \\ \midrule
InternVL2.5-78B & \multicolumn{3}{c}{In-context} & 0.25 & 0.30 & 0.46\\
\midrule
\multirow{4}{*}{LLaVA-NeXT-7B}  & - & - & - & 0.36 & 0.33 & 0.49\\
& \checkmark & - & - & 0.43 & 0.35 & 0.53 \\
& \checkmark & \checkmark & - & 0.48 & 0.37 & 0.54 \\
& \checkmark & \checkmark & \checkmark & \textbf{0.56} & \textbf{0.49} & \textbf{0.74}\\ 
\midrule
\multirow{4}{*}{Qwen2-VL-7B}  & - & -  & - & 0.39 & 0.38 & 0.53\\
& \checkmark & - & - & 0.48 & 0.40 & 0.56\\
& \checkmark & \checkmark & - & 0.50 & 0.42 & 0.56\\
& \checkmark & \checkmark & \checkmark & \textbf{0.61} & \textbf{0.50} & \textbf{0.85}\\ 
\midrule 
\multirow{4}{*}{InternVL2.5-8B}  & - & - & - & 0.42 & 0.45 & 0.71\\ 
& \checkmark & - & - &  0.52 & 0.48 & 0.61\\ 
& \checkmark & \checkmark & - & 0.53 & 0.50 & 0.62\\
& \checkmark & \checkmark & \checkmark & \textbf{0.63} & \textbf{0.52} & \textbf{0.82}\\ 
\bottomrule
\end{tabular}
\end{table}

\subsubsection{Evaluation Metrics}\label{sec:evaluation-metrics}

We assess the quality of each generated \emph{title\,+\,aspect}--JSON string using one holistic metric and two targeted metrics that focus on aspect-level correctness. These collectively evaluate both the fluency of the structured output and its alignment with e-commerce expectations.

\noindent\paragraph{\textbf{ROUGE--L F1.}}
To evaluate the overall quality of the generated JSON string as a complete entity, we use ROUGE--L F1, which captures both contextual coherence and structural fidelity. Given the predicted token sequence $s^{\mathrm{pred}}$ and the reference $s^{\mathrm{ref}}$, we first compute the length $l$ of their longest common subsequence (LCS). Then:
\[
P_{\mathrm{LCS}}=\frac{l}{|s^{\mathrm{pred}}|},\quad
R_{\mathrm{LCS}}=\frac{l}{|s^{\mathrm{ref}}|},\quad
\text{ROUGE--L\,F1}=\frac{2P_{\mathrm{LCS}}R_{\mathrm{LCS}}}
                           {P_{\mathrm{LCS}}+R_{\mathrm{LCS}}}\, .
\]
Since the entire structured string is compared, this metric rewards both semantic accuracy and adherence to the expected format.

\noindent\paragraph{\textbf{Aspect-Matching F1.}}
To further evaluate the model’s performance on aspect prediction, we extract and normalize all aspect name–value pairs from both the prediction and the reference. Let $\mathcal{A}^{\mathrm{pred}}$ and $\mathcal{A}^{\mathrm{ref}}$ denote the predicted and reference aspect sets, respectively. We define:
\[
P_{\text{asp}}
=\frac{|\mathcal{A}^{\mathrm{pred}}\cap\mathcal{A}^{\mathrm{ref}}|}
       {|\mathcal{A}^{\mathrm{pred}}|},\quad
R_{\text{asp}}
=\frac{|\mathcal{A}^{\mathrm{pred}}\cap\mathcal{A}^{\mathrm{ref}}|}
       {|\mathcal{A}^{\mathrm{ref}}|},
\]
\[
\text{Aspect F1}
=\frac{2P_{\text{asp}}R_{\text{asp}}}{P_{\text{asp}}+R_{\text{asp}}}\, .
\]
This fine-grained metric reflects how accurately the model identifies and reproduces key attribute pairs, which is central to downstream applications.

\noindent\paragraph{\textbf{Schema-Compliance Recall.}}
Finally, we evaluate whether the predicted aspect keys conform to the platform’s predefined schema. This is critical for ensuring that the outputs can be directly consumed by the e-commerce backend. Let $\mathcal{S}$ be the set of allowed schema keys and $K^{\mathrm{pred}}$ the predicted keys. Then:
\[
\text{Schema}_{\text{Recall}} = \frac{|\{ k \in K^{\mathrm{pred}} \mid k \in \mathcal{S} \}|}{|K^{\mathrm{pred}}|}.
\]
This metric reports the proportion of predicted keys that align with the schema, reflecting practical deployability in product listings.

\subsubsection{Effectiveness of OPAL Generalizes Across MLLM Backbones}  
\label{qualititive}
We evaluated OPAL using different MLLM backbones, LLaVA-NeXT-7B \cite{liu2024llavanext}, Qwen2-VL-7B \cite{wang2024qwen2}, and InternVL2.5-8B \cite{chen2024expanding}, on four variants: \textbf{(1) Baseline:} Raw product knowledge without conformity alignment, \textbf{(2) MACE:} Data refined to align textual and visual information, \textbf{(3) MACE + LACU:} Further enriched data capturing detailed product context, and \textbf{(4) MACE + LACU + DPO:} Enhanced contextual understanding through Direct Preference Optimization. As shown in \Cref{tab:main_results}, the proposed methods collectively improved generation quality, enhanced contextual understanding, and optimized overall performance across all backbones. Specifically, models trained with the full OPAL pipeline (MACE + LACU + DPO) achieve at least 50\% improvement in Rouge-L F1 (0.63 vs.\ 0.42) and at least 16\% improvement in Aspect F1 (0.52 vs.\ 0.45) compared to models trained on unaligned data. While higher scores consistently reflect improved generation quality, we do not expect metrics like Rouge-L F1 or Aspect F1 to reach 1.0, as many items can have multiple valid titles and aspect combinations. However, in this context, improvements from moderate to higher scores are still meaningful. Even though fine-tuned InternVL2.5-8B, using unaligned data, shows good Schema Recall (aspect name), the low Rouge F1 and Aspect F1 demonstrate significant hallucination in the aspect value.  We also compared with the common industrial practice, in-context learning, using a much larger backbone (InternVL2.5-78B), using category-specific demonstration examples. Despite the larger model size, in-context learning showed significantly lower generation quality, underscoring the advantages of OPAL’s fine-tuning approach for improving contextual understanding.

\subsection{Qualitative Evaluation}
\subsubsection{Effectiveness of the MACE and LACU Pipeline.} 
\begin{figure}[h]
    \centering
    \includegraphics[width=0.5\textwidth]{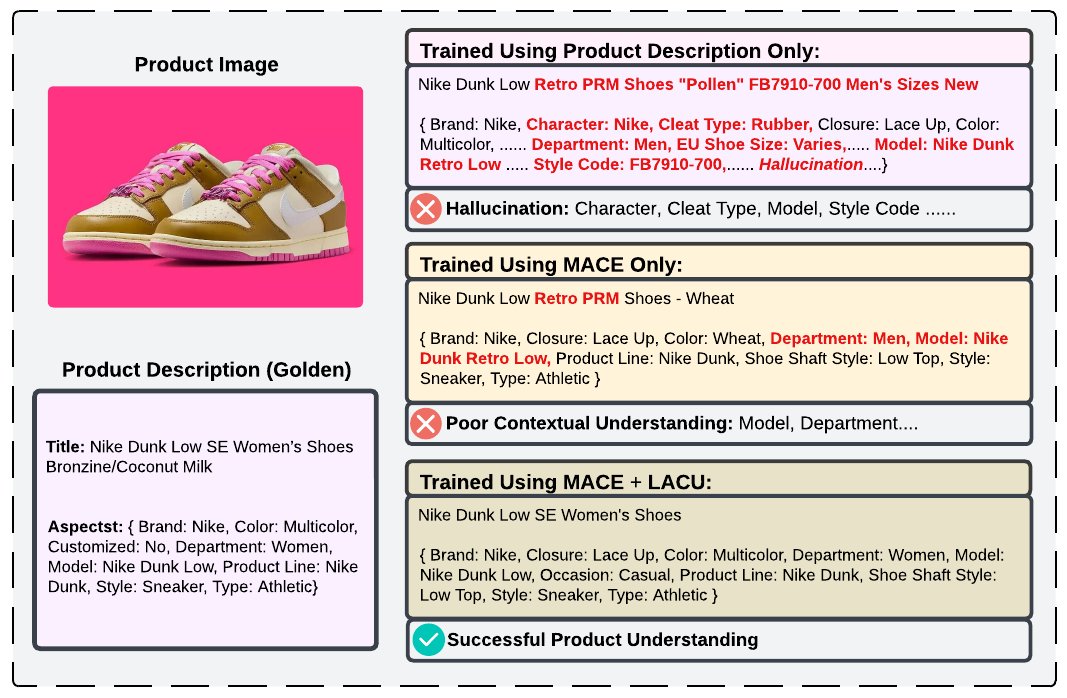}
    \caption{Qualitative comparison of the MLLM (InternVL2.5-8B) generated from different stages of the training (hallucination is highlighted in red).}
    \label{fig:fig3}
\end{figure}

We qualitatively illustrate the effectiveness of the MACE and LACU pipeline by visualizing generation outputs at different stages of the data refinement process in Figure~\ref{fig:fig3}. We evaluate an MLLM (InternVL2.5-8B) trained under varying data preparation regimes, paralleling the quantitative trends reported in Table~\ref{tab:main_results}. When trained on raw, unaligned descriptions (Baseline in Sec.\ref{qualititive}), the model produces a significant number of hallucinated attributes, limiting its applicability in real-world settings. While applying MACE helps reduce hallucinations by aligning image-text pairs (MACE), the model remains limited in extracting fine-grained contextual cues, such as specific shoe models, which are often underrepresented or missing in pretraining corpora. Upon further incorporating visual-grounded Q\&A supervision through LACU (MACE + LACU), the model demonstrates enhanced contextual comprehension and more accurate visual attribute generation. The results underscore the effectiveness of both MACE and LACU.

\subsubsection{OPAL Result Visualizations.}

\begin{table}[t]
\centering
\caption{Model outputs from the OPAL framework using InternVL2.5-8B. Each row shows the input image, generated title, and predicted aspects, highlighting the model’s ability to interpret niche items and seller intent.}
\label{tab:image_model_output}
\renewcommand{\arraystretch}{1.2}
\footnotesize
\begin{tabular}{|c|m{2cm}|m{5.4cm}|}
\hline
\textbf{\#} & \textbf{Image} & \textbf{Model Output (Title + Aspects)} \\
\hline

1 & 
\centering\includegraphics[width=2cm, keepaspectratio]{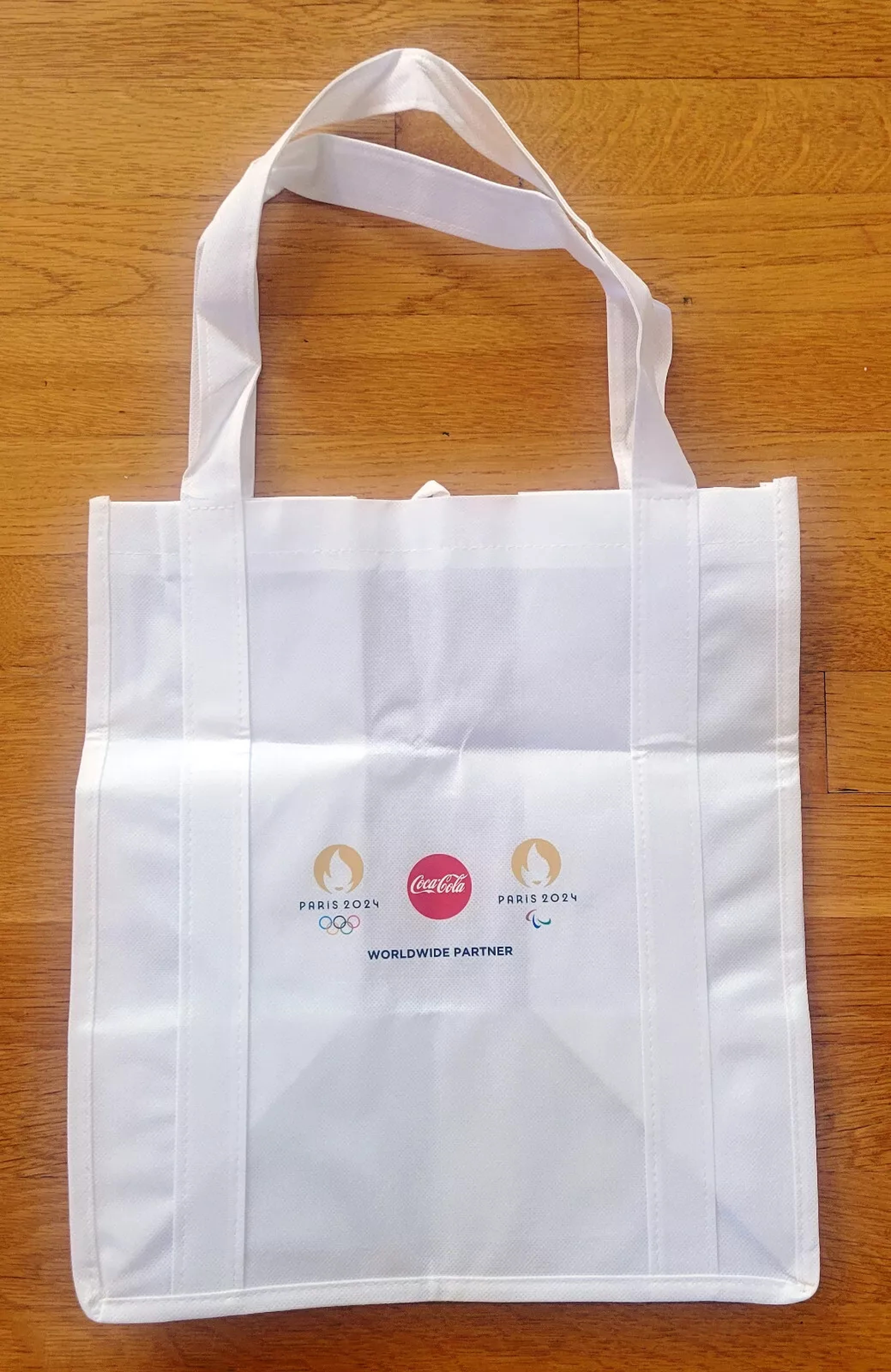} &
\textbf{Title:} Coca-Cola 2024 Paris Summer Olympics Tote Bag

\vspace{3pt}

\textbf{Aspects:} \{Brand: Coca-Cola, Color: White, Gender: Unisex Adult, Modified Item: No, Officially Licensed: Yes, Product: Tote Bag, Sport: Soccer, Team: Olympics\}
\\
\hline

2 & 
\centering\includegraphics[width=2cm, keepaspectratio]{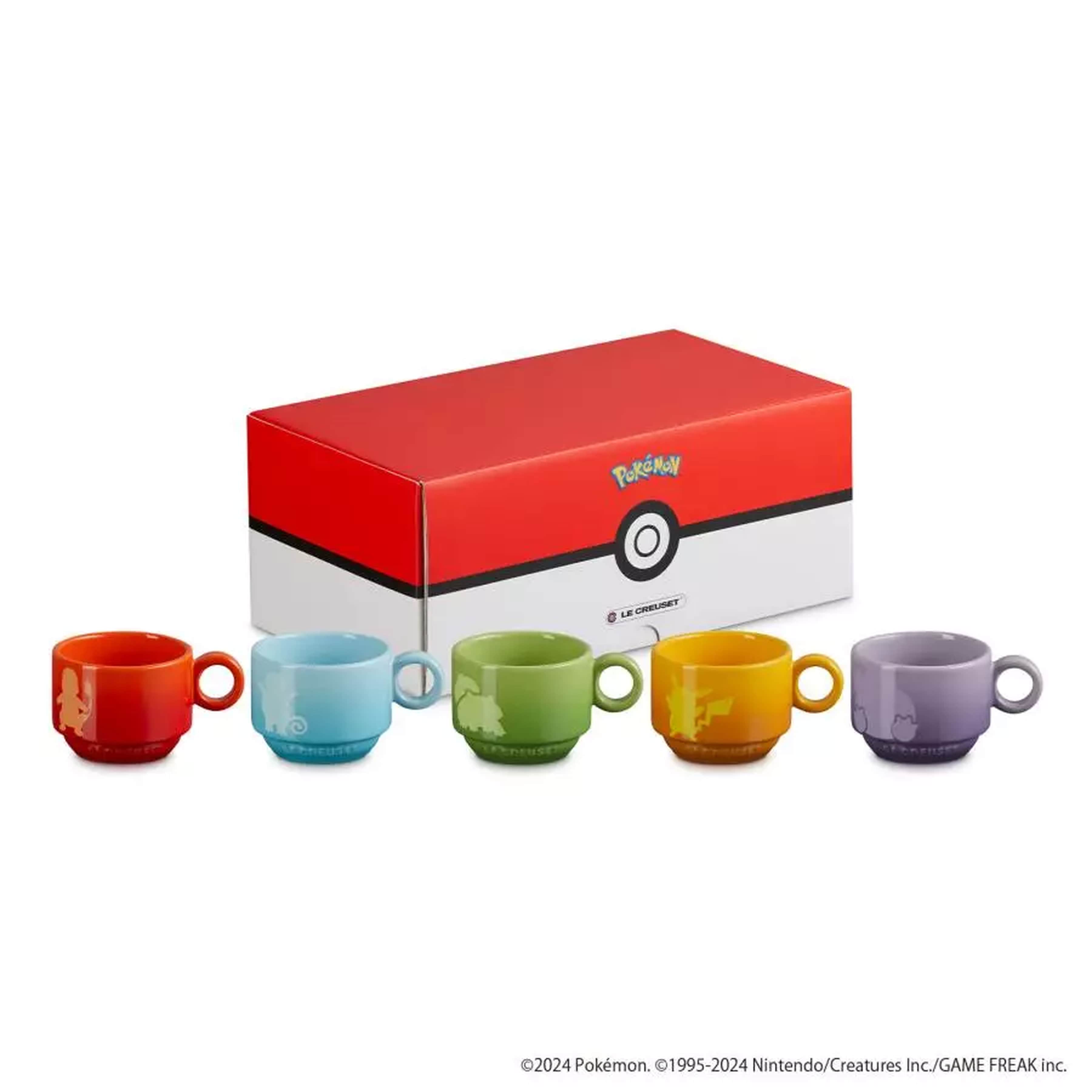} &
\textbf{Title:} Pokemon x Le Creuset Pokemon Special Edition Coffee Mug Set

\vspace{3pt}

\textbf{Aspects:} \{Brand: Le Creuset, Type: Coffee Mug Set, Theme: Pokemon, Material: Ceramic, Color: Multicolor, Number of Items in Set: Five-Piece, Features: Dishwasher Safe\}
\\
\hline

3 & 
\centering\includegraphics[width=2cm, keepaspectratio]{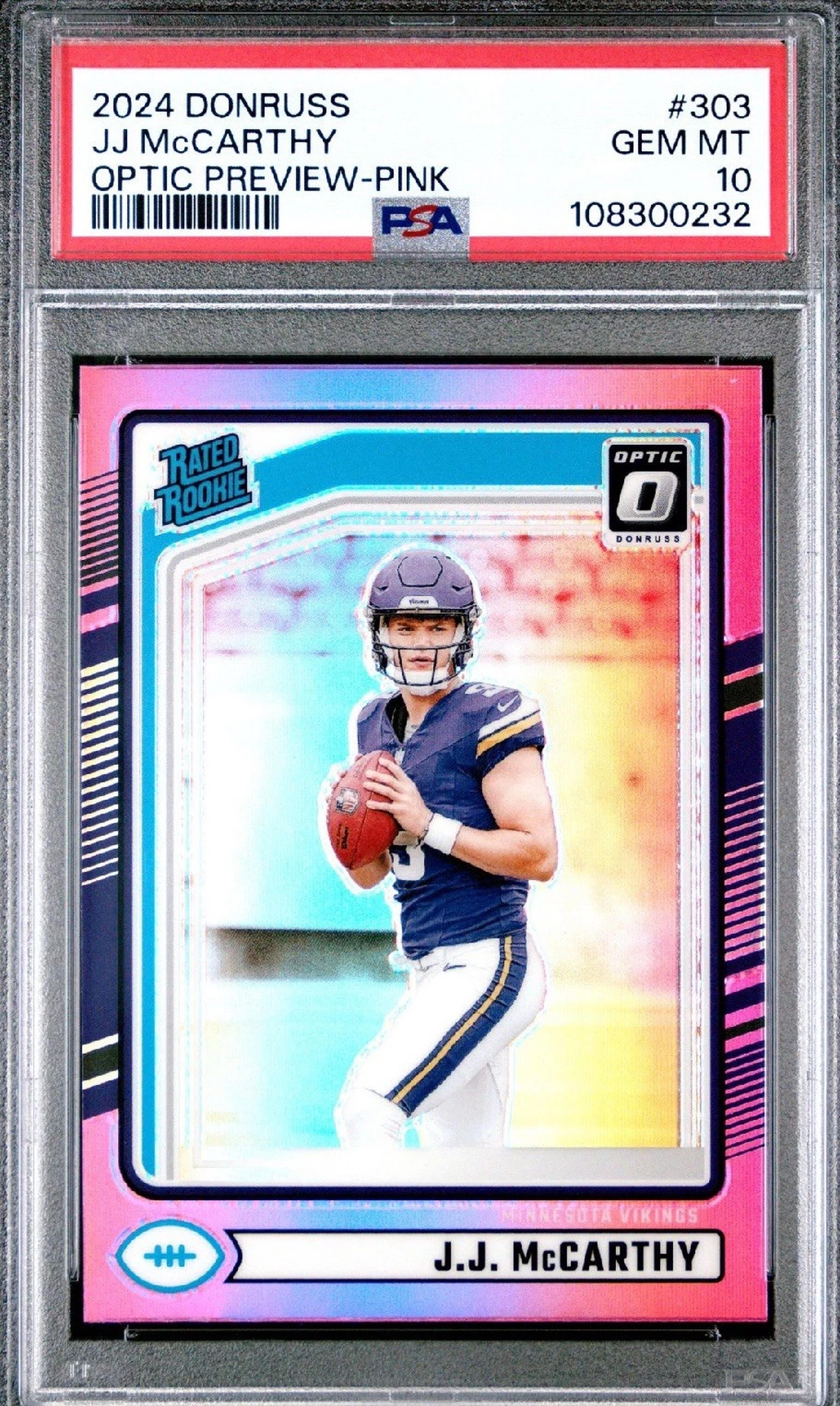} &
\textbf{Title:} 2024 Donruss Optic Preview Pink Pulsar Rated Rookie RC \#303 J.J. McCarthy

\vspace{3pt}

\textbf{Aspects:} \{Features: Rookie, Manufacturer: Panini, Parallel/Variety: Pink, Season: 2024, Set: 2024 Donruss Optic, Sport: Football, Type: Sports Trading Card, Player: J.J. McCarthy, Team: Minnesota Vikings, Card Number: 303, Grade: PSA 10 GEM MT, PSA Certification Number: 108300232\}
\\
\hline

4 & 
\centering\includegraphics[width=2cm, keepaspectratio]{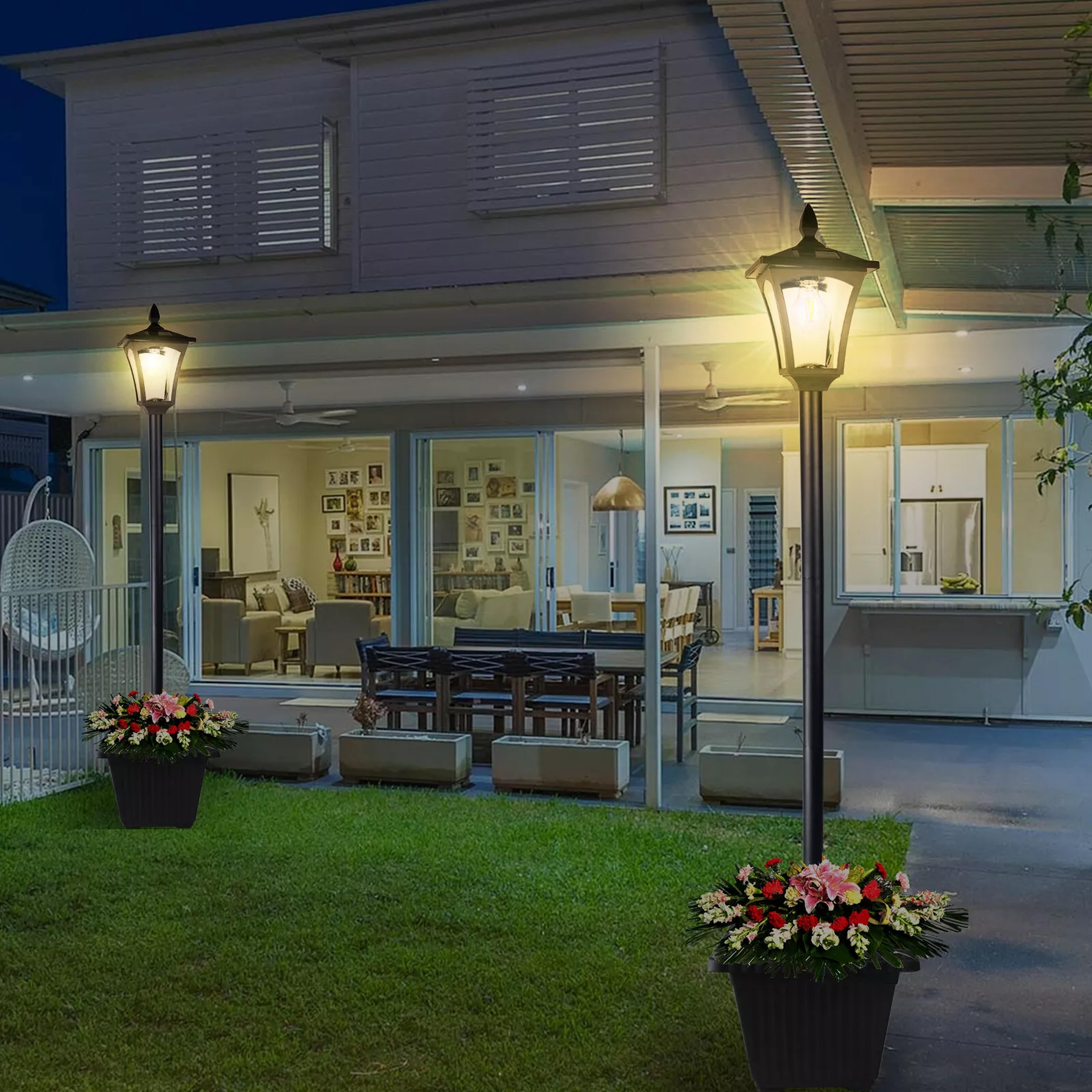} &
\textbf{Title:} Solar LED Landscape Pathway Lights, 2 Lights, Dusk to Dawn, Waterproof, Garden

\vspace{3pt}

\textbf{Aspects:} \{Brand: Unbranded, Bulb Type: LED, Features: Dusk-to-Dawn, Waterproof, Power Source: Solar, Style: Art Deco, Type: Deck/Step Light\}
\\
\hline

5 & 
\centering\includegraphics[width=2cm, keepaspectratio]{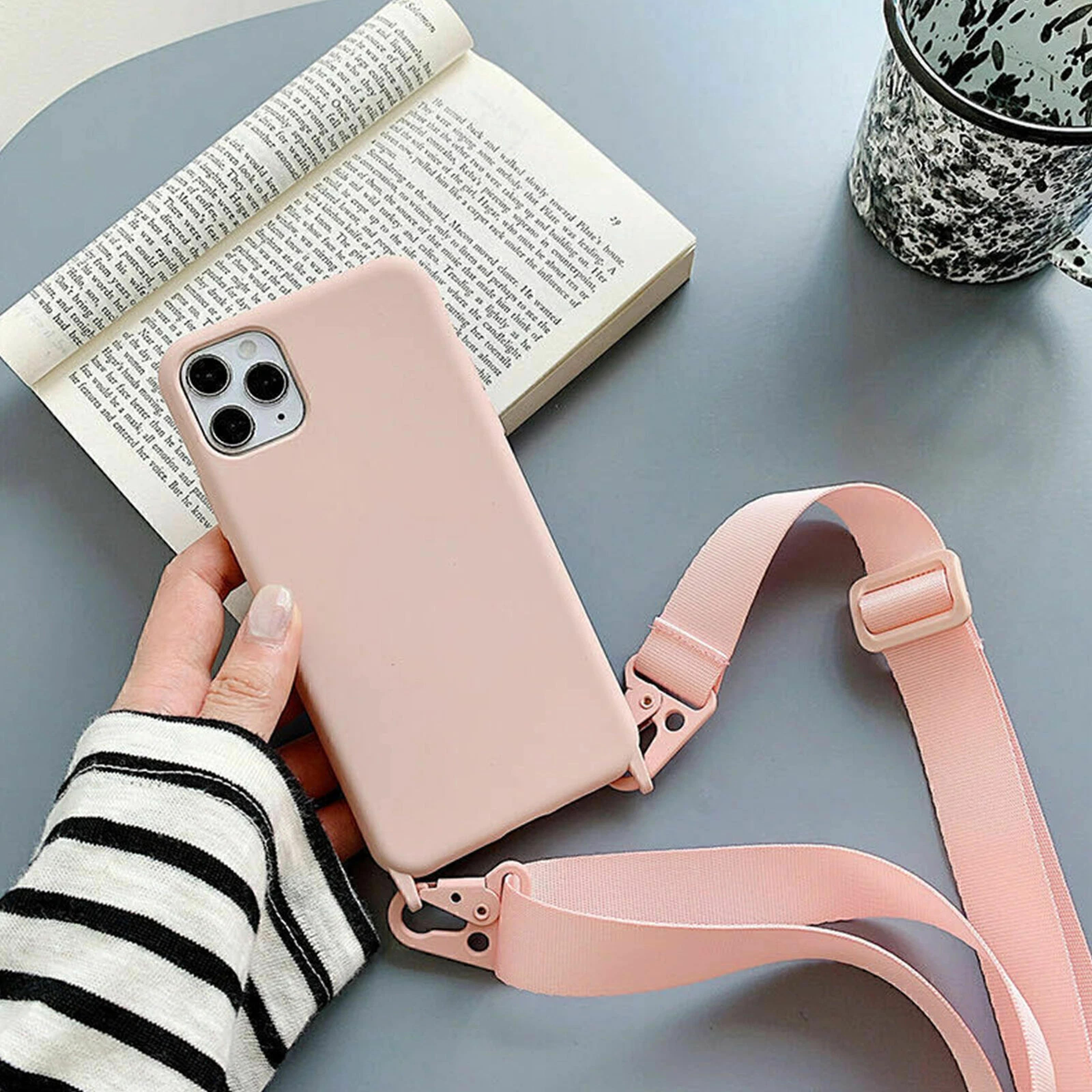} &
\textbf{Title:} Soft Silicone Phone Case Cover with Lanyard

\vspace{3pt}

\textbf{Aspects:} \{Brand: Unbranded, Compatible Brand: For Apple, Design/Finish: Luxury, Features: Shockproof, Material: Silicone/Gel/Rubber, Type: Back Case\}
\\
\hline

\end{tabular}
\end{table}

We further demonstrate the generations from the OPAL framework (InternVL2.5-8B) in Table~\ref{tab:image_model_output}. These cases highlight the model’s ability to accurately infer seller intent, extract fine-grained product aspects, and generalize to niche items that fall outside its training distribution. Notably, the first three examples were released after April 2024 and were absent from the training data, which was limited to sources available prior to March 2024. As a result, image similarity-based retrieval methods would be ineffective, given the lack of indexed product knowledge. 

In the first example, the key identifying features, namely the Coca-Cola and Paris Olympic icons, appear at the center of an otherwise predominantly white tote bag. Retrieval-based methods, even when the exact same bag is present in the inventory but photographed from slightly different angles or with varied shooting styles, tend to match such images to other generic white tote bags. This often results in incorrect product identification. The third example highlights OPAL’s ability to capture fine-grained details, such as the PSA certification number, card grade, and unique identifier. Although these elements occupy only a small portion of the image, they are critical and highly relevant to sellers. OPAL not only identifies these details but also infers contextual information, such as the card manufacturer and its parallel/variety, even when such information is not explicitly visible in the image. These capabilities demonstrate OPAL’s strong contextual reasoning and its potential to enhance product understanding in e-commerce. The last two examples demonstrate OPAL’s robustness in handling ambiguous images with limited visual cues regarding seller intent. Retrieval-based methods frequently misidentify these items by matching them to visually similar but semantically incorrect products. In contrast, OPAL exhibits strong domain-specific reasoning, generating accurate titles and aspect values that align with the true intent of the seller.

\begin{table}[]
\footnotesize
\centering
\caption{Offline comparison with baseline model}\label{tab:comparison_in_house}
\begin{tabular}{@{}c|c|c|c@{}}
\toprule
     & Final Submission Rate & Conformity & Win Rate \\ \midrule
OPAL (InternVL2.5-8B) & +53.40\%    & +10.50\%    & 60\%    \\ \bottomrule
\end{tabular}
\end{table}

\subsection{Comparison with a Baseline System}
To further evaluate the performance of OPAL, we compared it with a similarity-based baseline system leveraging multimodal retrieval. This pipeline employs a large-scale multimodal embedding model with a knowledge graph containing over one billion edges, meticulously developed over an 18-month period. The independent evaluation dataset comprises thousands of actual listing sessions, each featuring high-quality human annotations for titles and aspects. We employed three metrics for this comparison: \textbf{Final Submission Rate}: Assesses how effectively the model helps sellers complete listings when images are uploaded using an in-house user traffic simulation method. \textbf{Conformity}: Evaluated using a large MLLM (InternVL2.5-78B) to measure alignment with ground truth with alignment score from 1 to 5, similar as that in Table \ref{tab:dpo_prompt_example}, where 5 represents correct generation and 1 represents incorrect generation. \textbf{Win Rate}: Direct comparison with the baseline system, rated by the same large MLLM to determine if generation from the OPAL wins that from the baseline model. As demonstrated in \Cref{tab:comparison_in_house}, OPAL outperforms the baseline system by generating higher-quality titles and aspects, as well as significantly improving the final submission rate. These enhancements showcase a more simplified and efficient listing experience.

\section{Conclusion}
In this work, we introduced \textbf{O}ptimized \textbf{P}reference-Based \textbf{A}I for \textbf{L}istings (OPAL), a novel vision-language framework designed for e-commerce listing optimization. OPAL generates high-quality item information, including titles and structured aspects, directly from product images. It addresses key challenges in multimodal learning, such as the modality gap and lack of attribute-level context in LLM pretraining, through our proposed pipeline: (1) MACE filters and aligns noisy text with image-grounded data; (2) LACU injects contextual knowledge via visually grounded Q\&A generation; and (3) DPO fine-tunes outputs toward better contextual understanding. Experiments show OPAL outperforms strong baselines across multiple backbones and metrics, generalizing well to unseen or niche items. It effectively infers subtle product traits and seller intent, demonstrating strong contextual reasoning from multimodal data.

While alternative methods such as guided decoding using a very large MLLM or agent-based retrieval systems offer complementary value, they typically require extensive external infrastructure and are resource-intensive. Furthermore, they still fall short in bridging the underlying modality gap and contextual reasoning limitations of pretrained models. By contrast, OPAL offers a unified solution that embeds these capabilities directly through end-to-end fine-tuning, and it can still be incorporated into these advanced frameworks.

The OPAL is developed as an API to enable faster, comprehensive workflows for high-quality product information and schema-aligned product information from images. Beyond listing assistance, OPAL has broader applications across the e-commerce ecosystem. It can support a wide range of multimodal tasks, including image interpretation, product content understanding, and automated image labeling, serving as an internal tool to accelerate the annotation of training data for downstream models. Looking ahead, we plan to extend OPAL in several key directions. First, we aim to incorporate multi-image reasoning to aggregate information from diverse visual perspectives. Second, we can expand OPAL into an agentic framework capable of querying external knowledge, interacting with sellers, and adapting outputs based on category-specific constraints. These enhancements will move us closer to building an intelligent, end-to-end multimodal assistant for e-commerce.

Overall, this work underscores the transformative potential of fine-tuned vision-language models in streamlining and automating listing workflows, particularly for C2C platforms, where product metadata is often sparse or noisy. OPAL paves the way for more accurate, efficient, and user-aligned listing generation at scale.

\newpage

\appendix

\bibliography{acmart}


\begin{thebibliography}{39}


\ifx \showCODEN    \undefined \def \showCODEN     #1{\unskip}     \fi
\ifx \showDOI      \undefined \def \showDOI       #1{#1}\fi
\ifx \showISBNx    \undefined \def \showISBNx     #1{\unskip}     \fi
\ifx \showISBNxiii \undefined \def \showISBNxiii  #1{\unskip}     \fi
\ifx \showISSN     \undefined \def \showISSN      #1{\unskip}     \fi
\ifx \showLCCN     \undefined \def \showLCCN      #1{\unskip}     \fi
\ifx \shownote     \undefined \def \shownote      #1{#1}          \fi
\ifx \showarticletitle \undefined \def \showarticletitle #1{#1}   \fi
\ifx \showURL      \undefined \def \showURL       {\relax}        \fi
\providecommand\bibfield[2]{#2}
\providecommand\bibinfo[2]{#2}
\providecommand\natexlab[1]{#1}
\providecommand\showeprint[2][]{arXiv:#2}

\bibitem[Brinkmann et~al\mbox{.}(2025)]%
        {brinkmann2025extractgpt}
\bibfield{author}{\bibinfo{person}{Alexander Brinkmann}, \bibinfo{person}{Roee Shraga}, {and} \bibinfo{person}{Christian Bizer}.} \bibinfo{year}{2025}\natexlab{}.
\newblock \showarticletitle{ExtractGPT: Exploring the potential of Large Language Models for product attribute value extraction}. In \bibinfo{booktitle}{\emph{International Conference on Information Integration and Web Intelligence}}. Springer, \bibinfo{pages}{38--52}.
\newblock


\bibitem[Chen et~al\mbox{.}(2024b)]%
        {Chen2024Personalization}
\bibfield{author}{\bibinfo{person}{Jin Chen}, \bibinfo{person}{Zheng Liu}, \bibinfo{person}{Xu Huang}, \bibinfo{person}{Chenwang Wu}, \bibinfo{person}{Qi Liu}, \bibinfo{person}{Gangwei Jiang}, \bibinfo{person}{Yuanhao Pu}, \bibinfo{person}{Yuxuan Lei}, \bibinfo{person}{Xiaolong Chen}, \bibinfo{person}{Xingmei Wang}, \bibinfo{person}{Kai Zheng}, \bibinfo{person}{Defu Lian}, {and} \bibinfo{person}{Enhong Chen}.} \bibinfo{year}{2024}\natexlab{b}.
\newblock \showarticletitle{When large language models meet personalization: perspectives of challenges and opportunities}.
\newblock \bibinfo{journal}{\emph{World Wide Web}} \bibinfo{volume}{27}, \bibinfo{number}{4} (\bibinfo{year}{2024}), \bibinfo{pages}{42}.
\newblock


\bibitem[Chen et~al\mbox{.}(2024a)]%
        {chen2024internvl}
\bibfield{author}{\bibinfo{person}{Kang Chen} {et~al\mbox{.}}} \bibinfo{year}{2024}\natexlab{a}.
\newblock \showarticletitle{InternVL: Advancing Vision-Language Alignment with Large-Scale Pre-training}. In \bibinfo{booktitle}{\emph{Proceedings of the 2024 Conference on Empirical Methods in Natural Language Processing (EMNLP)}}.
\newblock


\bibitem[Chen et~al\mbox{.}(2024d)]%
        {Chen2024IPL}
\bibfield{author}{\bibinfo{person}{Kang Chen}, \bibinfo{person}{Qing Zhang}, \bibinfo{person}{Chengbao Lian}, \bibinfo{person}{Yixin Ji}, \bibinfo{person}{Xuwei Liu}, \bibinfo{person}{Shuguang Han}, \bibinfo{person}{Guoqiang Wu}, \bibinfo{person}{Fei Huang}, {and} \bibinfo{person}{Jufeng Chen}.} \bibinfo{year}{2024}\natexlab{d}.
\newblock \showarticletitle{IPL: Leveraging Multimodal Large Language Models for Intelligent Product Listing}. In \bibinfo{booktitle}{\emph{Proceedings of the 2024 Conference on Empirical Methods in Natural Language Processing: Industry Track}}. \bibinfo{publisher}{Association for Computational Linguistics}, \bibinfo{pages}{697--711}.
\newblock


\bibitem[Chen et~al\mbox{.}(2024c)]%
        {chen2024expanding}
\bibfield{author}{\bibinfo{person}{Zhe Chen}, \bibinfo{person}{Weiyun Wang}, \bibinfo{person}{Yue Cao}, \bibinfo{person}{Yangzhou Liu}, \bibinfo{person}{Zhangwei Gao}, \bibinfo{person}{Erfei Cui}, \bibinfo{person}{Jinguo Zhu}, \bibinfo{person}{Shenglong Ye}, \bibinfo{person}{Hao Tian}, \bibinfo{person}{Zhaoyang Liu}, {et~al\mbox{.}}} \bibinfo{year}{2024}\natexlab{c}.
\newblock \showarticletitle{Expanding Performance Boundaries of Open-Source Multimodal Models with Model, Data, and Test-Time Scaling}.
\newblock \bibinfo{journal}{\emph{arXiv preprint arXiv:2412.05271}} (\bibinfo{year}{2024}).
\newblock


\bibitem[Chowdhery et~al\mbox{.}(2022)]%
        {chowdhery2022palm}
\bibfield{author}{\bibinfo{person}{Aakanksha Chowdhery} {et~al\mbox{.}}} \bibinfo{year}{2022}\natexlab{}.
\newblock \showarticletitle{PaLM: Scaling Language Modeling with Pathways}.
\newblock \bibinfo{journal}{\emph{arXiv preprint arXiv:2204.02311}} (\bibinfo{year}{2022}).
\newblock


\bibitem[Devlin et~al\mbox{.}(2018)]%
        {devlin2018bert}
\bibfield{author}{\bibinfo{person}{Jacob Devlin}, \bibinfo{person}{Ming-Wei Chang}, \bibinfo{person}{Kenton Lee}, {and} \bibinfo{person}{Kristina Toutanova}.} \bibinfo{year}{2018}\natexlab{}.
\newblock \showarticletitle{BERT: Pre-training of Deep Bidirectional Transformers for Language Understanding}.
\newblock \bibinfo{journal}{\emph{arXiv preprint arXiv:1810.04805}} (\bibinfo{year}{2018}).
\newblock


\bibitem[Dubey et~al\mbox{.}(2024)]%
        {dubey2024llama}
\bibfield{author}{\bibinfo{person}{Abhimanyu Dubey}, \bibinfo{person}{Abhinav Jauhri}, \bibinfo{person}{Abhinav Pandey}, \bibinfo{person}{Abhishek Kadian}, \bibinfo{person}{Ahmad Al-Dahle}, \bibinfo{person}{Aiesha Letman}, \bibinfo{person}{Akhil Mathur}, \bibinfo{person}{Alan Schelten}, \bibinfo{person}{Amy Yang}, \bibinfo{person}{Angela Fan}, {et~al\mbox{.}}} \bibinfo{year}{2024}\natexlab{}.
\newblock \showarticletitle{The llama 3 herd of models}.
\newblock \bibinfo{journal}{\emph{arXiv preprint arXiv:2407.21783}} (\bibinfo{year}{2024}).
\newblock


\bibitem[Herold et~al\mbox{.}(2025)]%
        {herold2025domain}
\bibfield{author}{\bibinfo{person}{Christian Herold}, \bibinfo{person}{Michael Kozielski}, \bibinfo{person}{Tala Bazazo}, \bibinfo{person}{Pavel Petrushkov}, \bibinfo{person}{Seyyed~Hadi Hashemi}, \bibinfo{person}{Patrycja Cieplicka}, \bibinfo{person}{Dominika Basaj}, {and} \bibinfo{person}{Shahram Khadivi}.} \bibinfo{year}{2025}\natexlab{}.
\newblock \showarticletitle{Domain Adaptation of Foundation LLMs for e-Commerce}.
\newblock \bibinfo{journal}{\emph{arXiv preprint arXiv:2501.09706}} (\bibinfo{year}{2025}).
\newblock


\bibitem[Hongwimol et~al\mbox{.}(2025)]%
        {hongwimol2025gavel}
\bibfield{author}{\bibinfo{person}{Pollawat Hongwimol}, \bibinfo{person}{Dong Sheng}, \bibinfo{person}{Li Zhang}, \bibinfo{person}{Kai Liu}, {and} \bibinfo{person}{Xiufei Wang}.} \bibinfo{year}{2025}\natexlab{}.
\newblock \showarticletitle{GAVEL: Generative Attribute-Value Extraction Using LLMs on LLM-Augmented Datasets}. In \bibinfo{booktitle}{\emph{Proceedings of the 4th International Workshop on Knowledge-Augmented Methods for Natural Language Processing}}. \bibinfo{pages}{81--90}.
\newblock


\bibitem[Hu et~al\mbox{.}(2022)]%
        {jia2022scaling}
\bibfield{author}{\bibinfo{person}{Xiaowei Hu}, \bibinfo{person}{Zhe Gan}, \bibinfo{person}{Jianfeng Wang}, \bibinfo{person}{Zhengyuan Yang}, \bibinfo{person}{Zicheng Liu}, \bibinfo{person}{Yumao Lu}, {and} \bibinfo{person}{Lijuan Wang}.} \bibinfo{year}{2022}\natexlab{}.
\newblock \showarticletitle{Scaling Up Vision-Language Pre-Training for Image Captioning}. In \bibinfo{booktitle}{\emph{Proceedings of the IEEE/CVF Conference on Computer Vision and Pattern Recognition (CVPR)}}. \bibinfo{pages}{17980--17989}.
\newblock


\bibitem[Hu et~al\mbox{.}(2024)]%
        {Hu2024Denoised}
\bibfield{author}{\bibinfo{person}{Zhizhang Hu}, \bibinfo{person}{Shasha Li}, \bibinfo{person}{Ming Du}, \bibinfo{person}{Arnab Dhua}, {and} \bibinfo{person}{Douglas Gray}.} \bibinfo{year}{2024}\natexlab{}.
\newblock \showarticletitle{De-noised Vision-language Fusion Guided by Visual Cues for E-commerce Product Search}. In \bibinfo{booktitle}{\emph{IEEE/CVF Conference on Computer Vision and Pattern Recognition - Workshops}}. \bibinfo{publisher}{IEEE}, \bibinfo{pages}{1986--1996}.
\newblock


\bibitem[Kirillov et~al\mbox{.}(2023)]%
        {Kirillov_2023_ICCV}
\bibfield{author}{\bibinfo{person}{Alexander Kirillov}, \bibinfo{person}{Eric Mintun}, \bibinfo{person}{Nikhila Ravi}, \bibinfo{person}{Hanzi Mao}, \bibinfo{person}{Chloe Rolland}, \bibinfo{person}{Laura Gustafson}, \bibinfo{person}{Tete Xiao}, \bibinfo{person}{Spencer Whitehead}, \bibinfo{person}{Alexander~C. Berg}, \bibinfo{person}{Wan-Yen Lo}, \bibinfo{person}{Piotr Dollar}, {and} \bibinfo{person}{Ross Girshick}.} \bibinfo{year}{2023}\natexlab{}.
\newblock \showarticletitle{Segment Anything}. In \bibinfo{booktitle}{\emph{Proceedings of the IEEE/CVF International Conference on Computer Vision (ICCV)}}. \bibinfo{pages}{4015--4026}.
\newblock


\bibitem[Kuznetsova et~al\mbox{.}(2020)]%
        {kuznetsova2020open}
\bibfield{author}{\bibinfo{person}{Alina Kuznetsova}, \bibinfo{person}{Hassan Rom}, \bibinfo{person}{Neil Alldrin}, \bibinfo{person}{Jasper Uijlings}, \bibinfo{person}{Ivan Krasin}, \bibinfo{person}{Jordi Pont-Tuset}, \bibinfo{person}{Shahab Kamali}, \bibinfo{person}{Stefan Popov}, \bibinfo{person}{Matteo Malloci}, \bibinfo{person}{Alexander Kolesnikov}, {et~al\mbox{.}}} \bibinfo{year}{2020}\natexlab{}.
\newblock \showarticletitle{The open images dataset v4: Unified image classification, object detection, and visual relationship detection at scale}.
\newblock \bibinfo{journal}{\emph{International journal of computer vision}} \bibinfo{volume}{128}, \bibinfo{number}{7} (\bibinfo{year}{2020}), \bibinfo{pages}{1956--1981}.
\newblock


\bibitem[Li et~al\mbox{.}(2021)]%
        {Li2021Personalized}
\bibfield{author}{\bibinfo{person}{Lei Li}, \bibinfo{person}{Yongfeng Zhang}, {and} \bibinfo{person}{Li Chen}.} \bibinfo{year}{2021}\natexlab{}.
\newblock \showarticletitle{Personalized Transformer for Explainable Recommendation}. In \bibinfo{booktitle}{\emph{Proceedings of the 59th Annual Meeting of the Association for Computational Linguistics and the 11th International Joint Conference on Natural Language Processing}}. \bibinfo{publisher}{Association for Computational Linguistics}, \bibinfo{pages}{4947--4957}.
\newblock


\bibitem[Liang et~al\mbox{.}(2022)]%
        {ModalityGap}
\bibfield{author}{\bibinfo{person}{Weixin Liang}, \bibinfo{person}{Yuhui Zhang}, \bibinfo{person}{Yongchan Kwon}, \bibinfo{person}{Serena Yeung}, {and} \bibinfo{person}{James Zou}.} \bibinfo{year}{2022}\natexlab{}.
\newblock \showarticletitle{Mind the Gap: Understanding the Modality Gap in Multi-modal Contrastive Representation Learning}. In \bibinfo{booktitle}{\emph{NeurIPS}}.
\newblock
\urldef\tempurl%
\url{https://openreview.net/forum?id=S7Evzt9uit3}
\showURL{%
\tempurl}


\bibitem[Lin et~al\mbox{.}(2014)]%
        {lin2014microsoft}
\bibfield{author}{\bibinfo{person}{Tsung-Yi Lin}, \bibinfo{person}{Michael Maire}, \bibinfo{person}{Serge Belongie}, \bibinfo{person}{James Hays}, \bibinfo{person}{Pietro Perona}, \bibinfo{person}{Deva Ramanan}, \bibinfo{person}{Piotr Doll{\'a}r}, {and} \bibinfo{person}{C~Lawrence Zitnick}.} \bibinfo{year}{2014}\natexlab{}.
\newblock \showarticletitle{Microsoft coco: Common objects in context}. In \bibinfo{booktitle}{\emph{Computer Vision--ECCV 2014: 13th European Conference, Zurich, Switzerland, September 6-12, 2014, Proceedings, Part V 13}}. Springer, \bibinfo{pages}{740--755}.
\newblock


\bibitem[Liu et~al\mbox{.}(2024)]%
        {liu2024llavanext}
\bibfield{author}{\bibinfo{person}{Haotian Liu}, \bibinfo{person}{Chunyuan Li}, \bibinfo{person}{Yuheng Li}, \bibinfo{person}{Bo Li}, \bibinfo{person}{Yuanhan Zhang}, \bibinfo{person}{Sheng Shen}, {and} \bibinfo{person}{Yong~Jae Lee}.} \bibinfo{year}{2024}\natexlab{}.
\newblock \bibinfo{title}{LLaVA-NeXT: Improved Reasoning, OCR, and World Knowledge}.
\newblock
\newblock


\bibitem[Liu et~al\mbox{.}(2023)]%
        {liu2023llava}
\bibfield{author}{\bibinfo{person}{Haotian Liu}, \bibinfo{person}{Chunyuan Li}, \bibinfo{person}{Qingyang Wu}, {and} \bibinfo{person}{Yong~Jae Lee}.} \bibinfo{year}{2023}\natexlab{}.
\newblock \showarticletitle{Visual Instruction Tuning}. In \bibinfo{booktitle}{\emph{Advances in Neural Information Processing Systems}}, Vol.~\bibinfo{volume}{36}. \bibinfo{publisher}{Curran Associates, Inc.}, \bibinfo{pages}{34892--34916}.
\newblock


\bibitem[Majumder et~al\mbox{.}(2020)]%
        {Majumder2020Representation}
\bibfield{author}{\bibinfo{person}{Bodhisattwa~Prasad Majumder}, \bibinfo{person}{Navneet Potti}, \bibinfo{person}{Sandeep Tata}, \bibinfo{person}{James~Bradley Wendt}, \bibinfo{person}{Qi Zhao}, {and} \bibinfo{person}{Marc Najork}.} \bibinfo{year}{2020}\natexlab{}.
\newblock \showarticletitle{Representation Learning for Information Extraction from Form-like Documents}. In \bibinfo{booktitle}{\emph{Proceedings of the 58th Annual Meeting of the Association for Computational Linguistics}}. \bibinfo{publisher}{Association for Computational Linguistics}, \bibinfo{pages}{6495--6504}.
\newblock


\bibitem[Mandal et~al\mbox{.}(2023)]%
        {mandal2023semanticequivalenceecommercequeries}
\bibfield{author}{\bibinfo{person}{Aritra Mandal}, \bibinfo{person}{Daniel Tunkelang}, {and} \bibinfo{person}{Zhe Wu}.} \bibinfo{year}{2023}\natexlab{}.
\newblock \showarticletitle{Semantic Equivalence of e-Commerce Queries}.
\newblock
\showeprint[arxiv]{2308.03869}~[cs.IR]
\urldef\tempurl%
\url{https://arxiv.org/abs/2308.03869}
\showURL{%
\tempurl}


\bibitem[OpenAI(2023)]%
        {openai2023gpt}
\bibfield{author}{\bibinfo{person}{OpenAI}.} \bibinfo{year}{2023}\natexlab{}.
\newblock \showarticletitle{GPT-4 Technical Report}.
\newblock \bibinfo{journal}{\emph{arXiv preprint arXiv:2303.08774}} (\bibinfo{year}{2023}).
\newblock


\bibitem[Palen-Michel et~al\mbox{.}(2024)]%
        {Chester2024LLM}
\bibfield{author}{\bibinfo{person}{Chester Palen-Michel}, \bibinfo{person}{Ruixiang Wang}, \bibinfo{person}{Yipeng Zhang}, \bibinfo{person}{David Yu}, \bibinfo{person}{Canran Xu}, {and} \bibinfo{person}{Zhe Wu}.} \bibinfo{year}{2024}\natexlab{}.
\newblock \showarticletitle{Investigating {LLM} Applications in E-Commerce}.
\newblock \bibinfo{journal}{\emph{CoRR}}  \bibinfo{volume}{abs/2408.12779} (\bibinfo{year}{2024}).
\newblock


\bibitem[Peng et~al\mbox{.}(2024)]%
        {Peng2024ECeLLM}
\bibfield{author}{\bibinfo{person}{Bo Peng}, \bibinfo{person}{Xinyi Ling}, \bibinfo{person}{Ziru Chen}, \bibinfo{person}{Huan Sun}, {and} \bibinfo{person}{Xia Ning}.} \bibinfo{year}{2024}\natexlab{}.
\newblock \showarticletitle{eCeLLM: Generalizing Large Language Models for E-commerce from Large-scale, High-quality Instruction Data}. In \bibinfo{booktitle}{\emph{Forty-first International Conference on Machine Learning}}. \bibinfo{publisher}{OpenReview.net}.
\newblock


\bibitem[Plummer et~al\mbox{.}(2015)]%
        {plummer2015flickr30k}
\bibfield{author}{\bibinfo{person}{Bryan~A Plummer}, \bibinfo{person}{Liwei Wang}, \bibinfo{person}{Chris~M Cervantes}, \bibinfo{person}{Juan~C Caicedo}, \bibinfo{person}{Julia Hockenmaier}, {and} \bibinfo{person}{Svetlana Lazebnik}.} \bibinfo{year}{2015}\natexlab{}.
\newblock \showarticletitle{Flickr30k entities: Collecting region-to-phrase correspondences for richer image-to-sentence models}. In \bibinfo{booktitle}{\emph{Proceedings of the IEEE international conference on computer vision}}. \bibinfo{pages}{2641--2649}.
\newblock


\bibitem[Radford et~al\mbox{.}(2021)]%
        {radford2021learning}
\bibfield{author}{\bibinfo{person}{Alec Radford} {et~al\mbox{.}}} \bibinfo{year}{2021}\natexlab{}.
\newblock \showarticletitle{Learning Transferable Visual Models From Natural Language Supervision}. In \bibinfo{booktitle}{\emph{Proceedings of the International Conference on Machine Learning (ICML)}}.
\newblock


\bibitem[Rafailov et~al\mbox{.}(2024)]%
        {dpo}
\bibfield{author}{\bibinfo{person}{Rafael Rafailov}, \bibinfo{person}{Archit Sharma}, \bibinfo{person}{Eric Mitchell}, \bibinfo{person}{Stefano Ermon}, \bibinfo{person}{Christopher~D. Manning}, {and} \bibinfo{person}{Chelsea Finn}.} \bibinfo{year}{2024}\natexlab{}.
\newblock \showarticletitle{Direct preference optimization: your language model is secretly a reward model}. In \bibinfo{booktitle}{\emph{Proceedings of the 37th International Conference on Neural Information Processing Systems}} (New Orleans, LA, USA) \emph{(\bibinfo{series}{NIPS '23})}. \bibinfo{publisher}{Curran Associates Inc.}, \bibinfo{address}{Red Hook, NY, USA}, Article \bibinfo{articleno}{2338}, \bibinfo{numpages}{14}~pages.
\newblock


\bibitem[Rajbhandari et~al\mbox{.}(2020)]%
        {rajbhandari2020zero}
\bibfield{author}{\bibinfo{person}{Samyam Rajbhandari}, \bibinfo{person}{Jeff Rasley}, \bibinfo{person}{Olatunji Ruwase}, {and} \bibinfo{person}{Yuxiong He}.} \bibinfo{year}{2020}\natexlab{}.
\newblock \showarticletitle{Zero: Memory optimizations toward training trillion parameter models}. In \bibinfo{booktitle}{\emph{SC20: International Conference for High Performance Computing, Networking, Storage and Analysis}}. IEEE, \bibinfo{pages}{1--16}.
\newblock


\bibitem[Sharma et~al\mbox{.}(2018)]%
        {sharma2018conceptual}
\bibfield{author}{\bibinfo{person}{Piyush Sharma}, \bibinfo{person}{Nan Ding}, \bibinfo{person}{Sebastian Goodman}, {and} \bibinfo{person}{Radu Soricut}.} \bibinfo{year}{2018}\natexlab{}.
\newblock \showarticletitle{Conceptual captions: A cleaned, hypernymed, image alt-text dataset for automatic image captioning}. In \bibinfo{booktitle}{\emph{Proceedings of the 56th Annual Meeting of the Association for Computational Linguistics (Volume 1: Long Papers)}}. \bibinfo{pages}{2556--2565}.
\newblock


\bibitem[Touvron et~al\mbox{.}(2023)]%
        {touvron2023llama}
\bibfield{author}{\bibinfo{person}{Hugo Touvron} {et~al\mbox{.}}} \bibinfo{year}{2023}\natexlab{}.
\newblock \showarticletitle{LLaMA: Open and Efficient Foundation Language Models}.
\newblock \bibinfo{journal}{\emph{arXiv preprint arXiv:2307.09288}} (\bibinfo{year}{2023}).
\newblock


\bibitem[Wang et~al\mbox{.}(2024b)]%
        {wang2024yolov10}
\bibfield{author}{\bibinfo{person}{Ao Wang}, \bibinfo{person}{Hui Chen}, \bibinfo{person}{Lihao Liu}, \bibinfo{person}{Kai Chen}, \bibinfo{person}{Zijia Lin}, \bibinfo{person}{Jungong Han}, {and} \bibinfo{person}{Guiguang Ding}.} \bibinfo{year}{2024}\natexlab{b}.
\newblock \showarticletitle{Yolov10: Real-time end-to-end object detection}.
\newblock \bibinfo{journal}{\emph{arXiv preprint arXiv:2405.14458}} (\bibinfo{year}{2024}).
\newblock


\bibitem[Wang et~al\mbox{.}(2024a)]%
        {wang2024qwen2}
\bibfield{author}{\bibinfo{person}{Peng Wang}, \bibinfo{person}{Shuai Bai}, \bibinfo{person}{Sinan Tan}, \bibinfo{person}{Shijie Wang}, \bibinfo{person}{Zhihao Fan}, \bibinfo{person}{Jinze Bai}, \bibinfo{person}{Keqin Chen}, \bibinfo{person}{Xuejing Liu}, \bibinfo{person}{Jialin Wang}, \bibinfo{person}{Wenbin Ge}, {et~al\mbox{.}}} \bibinfo{year}{2024}\natexlab{a}.
\newblock \showarticletitle{Qwen2-vl: Enhancing vision-language model's perception of the world at any resolution}.
\newblock \bibinfo{journal}{\emph{arXiv preprint arXiv:2409.12191}} (\bibinfo{year}{2024}).
\newblock


\bibitem[Xu et~al\mbox{.}(2021)]%
        {xu-etal-2021-videoclip}
\bibfield{author}{\bibinfo{person}{Hu Xu}, \bibinfo{person}{Gargi Ghosh}, \bibinfo{person}{Po-Yao Huang}, \bibinfo{person}{Dmytro Okhonko}, \bibinfo{person}{Armen Aghajanyan}, \bibinfo{person}{Florian Metze}, \bibinfo{person}{Luke Zettlemoyer}, {and} \bibinfo{person}{Christoph Feichtenhofer}.} \bibinfo{year}{2021}\natexlab{}.
\newblock \showarticletitle{{V}ideo{CLIP}: Contrastive Pre-training for Zero-shot Video-Text Understanding}. In \bibinfo{booktitle}{\emph{Proceedings of the 2021 Conference on Empirical Methods in Natural Language Processing}}, \bibfield{editor}{\bibinfo{person}{Marie-Francine Moens}, \bibinfo{person}{Xuanjing Huang}, \bibinfo{person}{Lucia Specia}, {and} \bibinfo{person}{Scott Wen-tau Yih}} (Eds.). \bibinfo{publisher}{Association for Computational Linguistics}, \bibinfo{address}{Online and Punta Cana, Dominican Republic}, \bibinfo{pages}{6787--6800}.
\newblock
\urldef\tempurl%
\url{https://doi.org/10.18653/v1/2021.emnlp-main.544}
\showDOI{\tempurl}


\bibitem[Xue et~al\mbox{.}(2023)]%
        {xue2023pumgpt}
\bibfield{author}{\bibinfo{person}{Wei Xue}, \bibinfo{person}{Zongyi Guo}, \bibinfo{person}{Baoliang Cui}, \bibinfo{person}{Zheng Xing}, \bibinfo{person}{Xiaoyi Zeng}, \bibinfo{person}{Xiufei Wang}, \bibinfo{person}{Shuhui Wu}, {and} \bibinfo{person}{Weiming Lu}.} \bibinfo{year}{2023}\natexlab{}.
\newblock \showarticletitle{PUMGPT: A Large Vision-Language Model for Product Understanding}.
\newblock \bibinfo{journal}{\emph{arXiv preprint arXiv:2308.09568}} (\bibinfo{year}{2023}).
\newblock


\bibitem[Zhang et~al\mbox{.}(2024)]%
        {Zhang2024LLAShoppingAssistant}
\bibfield{author}{\bibinfo{person}{Shuo Zhang}, \bibinfo{person}{Boci Peng}, \bibinfo{person}{Xinping Zhao}, \bibinfo{person}{Boren Hu}, \bibinfo{person}{Yun Zhu}, \bibinfo{person}{Yanjia Zeng}, {and} \bibinfo{person}{Xuming Hu}.} \bibinfo{year}{2024}\natexlab{}.
\newblock \showarticletitle{LLaSA: Large Language and E-Commerce Shopping Assistant}.
\newblock \bibinfo{journal}{\emph{CoRR}}  \bibinfo{volume}{abs/2408.02006} (\bibinfo{year}{2024}).
\newblock


\bibitem[Zhang et~al\mbox{.}(2022)]%
        {zhang2022contrastive}
\bibfield{author}{\bibinfo{person}{Yuhao Zhang}, \bibinfo{person}{Hang Jiang}, \bibinfo{person}{Yasuhide Miura}, \bibinfo{person}{Christopher~D Manning}, {and} \bibinfo{person}{Curtis~P Langlotz}.} \bibinfo{year}{2022}\natexlab{}.
\newblock \showarticletitle{Contrastive learning of medical visual representations from paired images and text}. In \bibinfo{booktitle}{\emph{Machine Learning for Healthcare Conference}}. PMLR, \bibinfo{pages}{2--25}.
\newblock


\bibitem[Zheng et~al\mbox{.}(2018)]%
        {zheng2018opentag}
\bibfield{author}{\bibinfo{person}{Jiajun Zheng} {et~al\mbox{.}}} \bibinfo{year}{2018}\natexlab{}.
\newblock \showarticletitle{OpenTag: Open Attribute Value Extraction from Product Titles and Descriptions}. In \bibinfo{booktitle}{\emph{Proceedings of the Conference on Empirical Methods in Natural Language Processing (EMNLP)}}.
\newblock


\bibitem[Zheng et~al\mbox{.}(2024)]%
        {zheng2024llamafactory}
\bibfield{author}{\bibinfo{person}{Yaowei Zheng}, \bibinfo{person}{Richong Zhang}, \bibinfo{person}{Junhao Zhang}, \bibinfo{person}{Yanhan Ye}, \bibinfo{person}{Zheyan Luo}, \bibinfo{person}{Zhangchi Feng}, {and} \bibinfo{person}{Yongqiang Ma}.} \bibinfo{year}{2024}\natexlab{}.
\newblock \showarticletitle{LlamaFactory: Unified Efficient Fine-Tuning of 100+ Language Models}. In \bibinfo{booktitle}{\emph{Proceedings of the 62nd Annual Meeting of the Association for Computational Linguistics (Volume 3: System Demonstrations)}}. \bibinfo{publisher}{Association for Computational Linguistics}, \bibinfo{address}{Bangkok, Thailand}.
\newblock
\urldef\tempurl%
\url{http://arxiv.org/abs/2403.13372}
\showURL{%
\tempurl}


\bibitem[Zhu et~al\mbox{.}(2023)]%
        {zhu2023vl}
\bibfield{author}{\bibinfo{person}{Jinguo Zhu}, \bibinfo{person}{Xiaohan Ding}, \bibinfo{person}{Yixiao Ge}, \bibinfo{person}{Yuying Ge}, \bibinfo{person}{Sijie Zhao}, \bibinfo{person}{Hengshuang Zhao}, \bibinfo{person}{Xiaohua Wang}, {and} \bibinfo{person}{Ying Shan}.} \bibinfo{year}{2023}\natexlab{}.
\newblock \showarticletitle{Vl-gpt: A generative pre-trained transformer for vision and language understanding and generation}.
\newblock \bibinfo{journal}{\emph{arXiv preprint arXiv:2312.09251}} (\bibinfo{year}{2023}).
\newblock


\end{thebibliography}
\bibliographystyle{ACM-Reference-Format}
\end{document}